
\hfuzz=5pt
\font\titlefont=cmbx10 scaled\magstep1
\def\llb{[\![}
\def\rrb{]\!]}

\magnification=1200

\null
\vskip 2cm
\centerline{\titlefont REPRESENTATIONS OF QUANTUM ALGEBRAS }
\smallskip
\centerline{\titlefont AND q-SPECIAL FUNCTIONS}
\vskip 2cm
\centerline{\bf Roberto Floreanini}
\smallskip
\centerline{Istituto Nazionale di Fisica Nucleare, Sezione di Trieste}
\centerline {Dipartimento di Fisica Teorica,
Universit\`a di Trieste}
\centerline{Strada Costiera 11, 34014 Trieste, Italy}
\vskip 1cm
\centerline{{\bf Luc Vinet}
\footnote{$^{(*)}$}{Supported in part by the National Sciences and
Engineering Research Council (NSERC) of Canada.}}
\smallskip
\centerline{Laboratoire de Physique Nucl\'eaire}
\centerline{and}
\centerline{Centre de Recherches Math\'ematiques}
\centerline{Universit\'e de Montr\'eal}
\centerline{Montr\'eal, Canada H3C 3J7}
\vskip 2cm
\centerline{\bf Abstract}
\smallskip\midinsert\narrower\narrower\noindent
The connection between $q$-analogs of special functions and representations
of quantum algebras has been developped recently. It has led to advances in the
theory of $q$-special functions that we here review.
\endinsert

\vfil\eject

\null
{\bf TABLE OF CONTENTS}
\bigskip
\item{1.} INTRODUCTION
\medskip
\item{2.} REVIEW OF q-ANALYSIS
\medskip
\item{3.} q-BESSEL FUNCTIONS AND THE TWO-DIMENSIONAL EUCLIDEAN QUANTUM ALGEBRA
\smallskip
\itemitem{3.1} One-variable realization and generating functions
\smallskip
\itemitem{3.2} Two-variable realization and addition formulas
\medskip
\item{4.} q-OSCILLATORS AND BASIC SPECIAL FUNCTIONS
\smallskip
\itemitem{4.1} One-variable model and $q$-Laguerre functions and polynomials
\smallskip
\itemitem{4.2} $q$-Oscillators and $q$-Hermite polynomials
\smallskip
\itemitem{4.3} The metaplectic representation of $su_q(1,1)$ and the
$q$-Gegenbauer polynomials
\medskip
\item{5.} THE QUANTUM ALGEBRA sl$_{\hbox{q}}$(2) AND q-HYPERGEOMETRIC FUNCTIONS
\smallskip
\itemitem{5.1} Realizations in terms of first-order $q$-difference operators
\smallskip
\itemitem{5.2} A realization in terms of second-order $q$-difference operators

\vskip 2cm

{\bf 1. INTRODUCTION}
\medskip

Most special functions of mathematical physics obey second-order differential
equations for which Lie algebra theory is well known to provide a natural
and useful setting.$^{1-3}$ Many of them have $q$-analogs$^4$ that satisfy
$q$-difference equations. The algebraic interpretation of these $q$-special
functions was initiated by Miller more than twenty years ago.$^5$ A lot of
interest in this subject developped recently when it was realized that
$q$-special functions are connected to quantum algebras and quantum
groups.$^{6,7}$ Advances in this direction$^{8-12}$ will be the subject
of the present review. (See also Refs.[13,14].)

We shall present realizations of quantized universal enveloping algebras
in terms of $q$-difference operators. In analogy with Lie theory,
we shall use the $q$-exponentials of the quantum algebra generators.
Various $q$-special functions will be identified as matrix elements
of these operators and also as basis vectors of the representation spaces.
This approach allows to derive many generating relations and addition formulas
that will be given.

In Section 2, we collect elements of $q$-analysis that will be used
throughout.

In Section 3, we discuss the two-dimensional Euclidean quantum algebra
${\cal E}_q(2)$. We establish its relation with $q$-Bessel
functions using a realization in one complex variable.
A two-variable realization of ${\cal E}_q(2)$ is also given from which
$q$-analogs of the Lommel and Graf addition formulas for the $q$-Bessel
functions are obtained.

Section 4 is devoted to $q$-oscillator representations. We first present
realizations of the oscillator quantum algebra on functions of one complex
variable. Irreducible modules on which the spectrum of the number operator $N$
is unbounded, bounded from below or above are examined. They provide an
interpretation of the $q$-Laguerre functions and polynomials.
An additional one-variable realization (with $N$ bounded below) is also given,
where the $q$-Hermite polynomials appear in the basis vectors. Finally,
the $q$-oscillators are used to construct the metaplectic representation
of $su_q(1,1)$ which gives an algebraic interpretation of a certain
$q$-generalization of the Gegenbauer polynomials.

In Section 5, we focus on $sl_q(2)$. We start by considering first-order
$q$-difference operators in one complex variable and discuss realizations
corresponding to representations that are unbounded, bounded from either below
or above and finite-dimensional. Basic hypergeometric functions of the type
$_2\phi_1$ are associated to the infinite-dimensional modules,
while little $q$-Jacobi polynomials arise in the finite-dimensional case.
We conclude by giving another realization of $sl_q(2)$, this time in terms of
second-order $q$-difference operators, which allows to obtain different
generating relations for the basic hypergeometric function $_2\phi_1$.

\vskip 2cm

{\bf 2. REVIEW OF q-ANALYSIS}
\medskip

We here collect results and formulas of $q$-analysis that will prove
useful.$^4$ For $a$ and $\alpha$ arbitrary complex numbers, we denote by
$(a;q)_\alpha$ the $q$-shifted factorial:
$$ (a;q)_\alpha={(a;q)_\infty\over (a q^\alpha;q)_\infty}\ ,\eqno(2.1)$$
where
$$ (a;q)_\infty=\, \prod_{k=0}^\infty (1-aq^k)\ ,\qquad\quad |q|<1\ .
\eqno(2.2)$$
Note that for $\alpha$ a positive integer $n$,
$$ (a;q)_n=(1-a)(1-aq)\dots (1-a q^{n-1})\ .\eqno(2.3)$$
These products satisfy various identities like for instance
$$ q^{n(n-1)\over2}\, (a^{-1} q^{1-n};q)_n=(-a^{-1})^n (a;q)_n\ .\eqno(2.4)$$
The $q$-binomial symbol is defined by
$$ \left[{n\atop m}\right]_q
={(q;q)_n\over (q;q)_m\, (q;q)_{n-m}}\ .\eqno(2.5)$$
Note that $(q;q)_n/(1-q)^n\rightarrow n!$ as $q\rightarrow 1^-$, so that (2.5)
reduces to the usual binomial symbol in that limit.
Of fundamental importance
is Heine's $q$-binomial theorem which states that
$$ \sum_{n=0}^\infty {(a;q)_n\over (q;q)_n} z^n={(az;q)_\infty\over
(z;q)_\infty}\ ,\qquad |z|<1,\ \ |q|<1\ .\eqno(2.6)$$
Two $q$-exponential functions are obtained from the above formula.
On the one hand, upon setting $a=0$, one gets
$$ e_q(z)=\sum_{n=0}^\infty {1\over (q;q)_n}\ z^n={1\over (z;q)_\infty}\ ,
\qquad |z|<1\ ,\eqno(2.7)$$
while on the other, upon replacing $z$ by $-z/a$ in (2.6), using (2.4) and
letting $a\rightarrow\infty$, one finds
$$ E_q(z)=\sum_{n=0}^\infty {q^{n(n-1)\over2}\over (q;q)_n}\ z^n=
(-z;q)_\infty\ .\eqno(2.8)$$
It is easy to see that $e_q(z)\, E_q(-z)=1$ and that
$$\lim_{q\rightarrow 1^-} e_q\bigl(z(1-q)\bigr)=
\lim_{q\rightarrow 1^-} E_q\bigl(z(1-q)\bigr)=\, e^z\ .\eqno(2.9)$$

Let $T_q$ be the $q$-dilatation operator acting as follows on functions
of the complex variable $z$
$$ T_q\, f(z)=\, f(qz)\ .\eqno(2.10)$$
The $q$-difference operators $D^+_z$ and $D^-_z$ are given by
$$ \eqalignno{ &D^+_z= z^{-1}(1-T_q)\ ,&(2.11a)\cr
               &D^-_z= z^{-1}(1-T_q^{-1})\ .&(2.11b)\cr}$$
Observe that ${1\over(1-q)} D^+_z\rightarrow d/dz$ and
${1\over(1-q^{-1})} D^-_z\rightarrow d/dz$ as $q\rightarrow 1$ and that
the $q$-exponentials obey
$$\eqalignno{ &D^+_z\, e_q(z)=\, e_q(z)\ ,&(2.12a)\cr
              &D^-_z\, E_q(z)=-q^{-1} E_q(z)\ .&(2.12b)\cr}$$

The basic hypergeometric series $_r\phi_s$ is defined by
$$\eqalign{ _r\phi_s (a_1,a_2,\ldots &,a_r;b_1,\ldots,b_s;q;z)\cr
&=\sum_{n=0}^\infty {(a_1;q)_n (a_2;q)_n\dots (a_r;q)_n\over
(q;q)_n (b_1;q)_n\dots (b_s;q)_n}\ \left[(-1)^n q^{n(n-1)\over2}
\right]^{1+s-r}\ z^n\ ,\cr} \eqno(2.13)$$
with $q\neq 0$ when $r>s+1$. Since $(q^{-m};q)_n=\, 0$, for
$n=m+1,m+2,\ldots$, the series $_r\phi_s$ terminates if one of the
numerator parameters $\{a_i\}$ is of the form $q^{-m}$ with $m=0,1,2\ldots$,
and $q\neq 0$.

When $b\rightarrow q^{-m}$, with $m$ a nonnegative integer,
the functions $_0\phi_1(b;q;z)$, $_1\phi_1(a;b;q;z)$ and
$_2\phi_1(a_1,a_2;b;q;z)$ satisfy the following limit relations:
$$_0\phi_1(q^{-m};q;z)\ (q^{-m};q)_{m+1}=
z^{m+1}\, {}_0\phi_1\bigl(q^{m+2};q;z\, q^{2(m+1)}\bigr)\
{q^{m(m+1)}\over (q;q)_{m+1}}\ ,\eqno(2.14)$$
\null
$$ _1\phi_1(a;q^{-m};z)\, (q^{-m};q)_{m+1}= z^{m+1}
{}_1\phi_1\bigl(a q^{m+1};q^{m+2};z q^{m+1}\bigr)\,(-1)^{m+1}\,
q^{{1\over2}m(m+1)}
{(a;q)_{m+1} \over (q;q)_{m+1}}\ ,\eqno(2.15)$$
\null
$$ {}_2\phi_1(a,b,q^{-m};q;z)\ (q^{-m};q)_{m+1}=z^{m+1}
{}_2\phi_1\bigl(a q^{m+1},b q^{m+1},q^{m+2};q;z\bigr)\
{(a;q)_{m+1} (b;q)_{m+1}\over (q;q)_{m+1}}\ .\eqno(2.16)$$
The last two formulas are the $q$-analogs of limit relations that the
confluent and ordinary hypergeometric functions, $_1F_1$ and $_2F_1$,
respectively, verify.$^{15,1}$

\vskip 2cm

{\bf 3. q-BESSEL FUNCTIONS AND THE TWO-DIMENSIONAL \hfill\break
\indent\phantom{3. }EUCLIDEAN QUANTUM ALGEBRA}
\bigskip
\indent\item{\bf 3.1}{\bf One-variable realization and generating functions}
\medskip

The Bessel functions of the first kind $J_\nu(z)$ have two
$q$-analogs:$^4$
$$\eqalignno{&J^{(1)}_\nu(z;q)={1\over (q;q)_\nu}\ \left({z\over2}\right)^\nu
\ {}_2\phi_1\Bigl(0,0;q^{\nu+1};q,-{z^2\over4}\Bigr)\ ,&(3.1a)\cr
&J_\nu^{(2)}(z;q)={1\over (q;q)_\nu}\, \left({z\over2}\right)^\nu\
{}_0\phi_1\Bigl(q^{\nu+1};q,-{z^2\, q^{\nu+1}\over4}\Bigr)\ ,&(3.1b)\cr}$$
for $0<q<1$. The function $J_\nu^{(2)}(z;q)$ is an entire transcendental
function and it is connected to $J_\nu^{(1)}(z;q)$ through the formula:
$$ J_\nu^{(2)}(z;q)=(-z^2/4;q)_\infty\ J_\nu^{(1)}(z;q)\ .\eqno(3.2)$$
One can check that in the limit $q\rightarrow 1^-$ both functions (3.1) reduce
to the ordinary Bessel functions,
$$\lim_{q\rightarrow 1^-}\,J_\nu^{(k)}\left((1-q)\,z;q\right)=
J_\nu(z)\ ,\qquad\quad k=1,2\ .\eqno(3.3)$$

The basic Bessel functions satisfy the following recursion relation:
$$ {(1-q^\nu)\over z}\, J_\nu^{(k)}(z;q)=
{1\over2}\left(J_{\nu-1}^{(k)}(z;q)+q^\nu\,J_{\nu+1}^{(k)}(z;q)\right)\ ,
\qquad k=1,2\ .\eqno(3.4)$$
Acting with the difference operator (2.11a) on $J_\nu^{(1)}(z;q)$
one can check that
$$ \left(D_z^+ -{z\over4}\right)\, J_\nu^{(1)}(z;q)=
{1\over2}\left(J_{\nu-1}^{(1)}(z;q)-J_{\nu+1}^{(1)}(z;q)\right)\ .\eqno(3.5)$$
An analogous formula holds for $J_\nu^{(2)}(z;q)$, where however the difference
operator $D_z^+$ is replaced by
$(-z^2/4;q)_\infty^{-1}\, D_z^+\, (-z^2/4;q)_\infty^{-1}$.
The relations (3.4) and (3.5) can be combined to give$^{12}$
$$\eqalignno{&\left[\Bigl(D_z^+ -{z\over4}\Bigr)\, q^\nu
+{(1-q^\nu)\over z}\right]\, J_\nu^{(1)}(z;q)={(1+q^\nu)\over2}\,
J_{\nu-1}^{(1)}(z;q)\ ,&(3.6a)\cr
&\left[-\Bigl(D_z^+ -{z\over4}\Bigr)
+{(1-q^\nu)\over z}\right]\, J_\nu^{(1)}(z;q)={(1+q^\nu)\over2}\,
J_{\nu+1}^{(1)}(z;q)\ .&(3.6b)\cr}$$

When the index $\nu$ is an integer $n$, with the help of the limit formula
(2.14) and of (3.2) one can check that:
$$ J^{(k)}_{-n}(z;q)=(-1)^n\ J^{(k)}_n(z;q)\ ,\qquad\quad k=1,2\ .\eqno(3.7)$$

The two-dimensional quantum algebra ${\cal E}_q(2)$ has the same defining
relations as its classical counterpart:
$$ [J,P_\pm]=\pm P_\pm\ ,\qquad\qquad [P_+,P_-]=\, 0\ .\eqno(3.8)$$
It is however endowed with a non-trivial Hopf structure
by taking the following definitions$^{16}$
of coproduct $\Delta$: ${\cal E}_q(2)\rightarrow$
$ {\cal E}_q(2)\otimes\, {\cal E}_q(2)$, antipode
$S$: ${\cal E}_q(2)\rightarrow {\cal E}_q(2)$ and counit
$\varepsilon$: ${\cal E}_q(2)\rightarrow {\bf C}$:
$$\eqalign{&\Delta(J)=J\otimes {\bf 1}+{\bf 1}\otimes J\ ,\phantom{q^J}\cr
           &S(J)=-J\ ,\phantom{q^J}\cr
           &\varepsilon(J)=\varepsilon(P_\pm)=\,0\ ,\cr}\qquad
  \eqalign{&\Delta(P_\pm)=P_\pm\otimes q^{-J/2}+q^{J/2}\otimes P_\pm\ ,\cr
           &S(P_\pm)=-q^{\mp 1/2}\, P_\pm\ ,\cr
           &\varepsilon({\bf 1})=1\ .\cr} \eqno(3.9)$$
As $q\rightarrow 1^-$ the trivial Hopf structure of ${\cal E}(2)$ is
recovered.

Take $J$ and $P_\pm$ to be the following operators acting on the space
$\cal H$ of all finite linear combinations of the functions $z^n$,
$z\in {\bf C}$, $n\in{\bf Z}$,
$$J=m_0+z{d\over dz}\ ,\qquad P_+=\omega\, z\ ,\qquad
P_-={\omega\over z}\ ;\eqno(3.10)$$
$m_0$ and $\omega$ are complex parameters, such that $\omega \neq 0$
and $0\leq {\Re e}\, m_0<1$. It is easily seen to give a representation
of the algebra (3.8), which we shall call $Q(\omega,m_0)$.
Define the basis vectors $f_m$ of $\cal H$ by $f_m(z)=z^n$, where
$m=m_0+n$ and $n\in {\bf Z}$. Then,
$$\eqalign{ &P_\pm \, f_m=\,\omega f_{m\pm 1}\ ,\cr
            &J\, f_m=m f_m\ .\cr}\eqno(3.11)$$
In analogy$^{13}$ with ordinary Lie theory, we introduce the operator
$$ U(\alpha, \beta,\gamma)=\, E_q\left(\alpha(1-q)P_+\right)\,
E_q\left(\beta(1-q)P_-\right)\, E_q\left(\gamma(1-q)J\right)\ ,\eqno(3.12)$$
which in the realization (3.10) becomes
$$ U(\alpha,\beta,\gamma)=\, E_q\bigl(\alpha\,\omega(1-q)z\bigr)\,
E_q\left(\beta\,\omega (1-q){1\over z}\right)\,
E_q\left(\gamma(1-q)\Bigl(m_0+z{d\over dz}\Bigr)\right)\ .\eqno(3.13)$$
We define the matrix elements $U_{kn}(\alpha,\beta,\gamma)$ through
$$ U(\alpha,\beta,\gamma)\ f_{m_0+n}=\sum_{k=-\infty}^\infty\,
U_{kn}(\alpha,\beta,\gamma)\ f_{m_0+k}\ .\eqno(3.14)$$

Using $(3.1b)$ and (2.14) the matrix elements $U_{kn}(\alpha,\beta,\gamma)$
are shown to have in general the following expression:$^{11}$
$$ \eqalign{U_{kn}(\alpha,\beta,\gamma)=E_q\bigl(\gamma\, (1-q)(m_0+n)\bigr)
&q^{{1\over2}(k-n)^2}\left(-{\alpha\over\beta}\right)^{(k-n)/2}\cr
&\qquad \times\ J^{(2)}_{k-n}\biggl(2\,\omega\,(1-q)
\left(-{\alpha\beta\over q}\right)^{1/2};q
\biggr)\ ;\cr}\eqno(3.15)$$
in the particular cases $\alpha=\gamma= 0$ and $\beta=\gamma= 0$,
the non-zero entries take the simpler form:
$$\eqalignno{&U_{kn}(0,\beta,0)=q^{{1\over2}(n-k)(n-k-1)}\,
{(\omega\beta(1-q))^{n-k}\over(q;q)_{n-k}}\ ,\qquad  k\leq n\ ,&(3.16a)\cr
             &U_{kn}(\alpha,0,0)=q^{{1\over2}(k-n)(k-n-1)}\,
{(\omega\alpha(1-q))^{k-n}\over(q;q)_{k-n}}\ ,\qquad  k\geq n\ .&(3.16b)\cr}$$

A generating function for the $q$-Bessel functions is now straightforwardly
obtained.  Setting $\alpha=1$, $\beta=-q$, $\gamma=\,0$, $n=\,0$ and
$2\,\omega\, (1-q)=x$, in (3.14) gives
$$ E_q\left({x\, z\over2}\right)\, E_q\left(-{q\, x\over 2 z}\right)=
\sum_{k=-\infty}^{\infty} q^{{1\over2}k(k-1)}\ J^{(2)}_k(x;q)\ z^k\ .
\eqno(3.17)$$
Similarly, using the $q$-exponential function $e_q$ defined in (2.7) instead of
$E_q$, one obtains the following generating function$^{4,5}$ for $J^{(1)}_k$
$$ e_q\left({x\, z\over2}\right)\, e_q\left(-{x\over 2 z}\right)=
\sum_{k=-\infty}^{\infty}  J^{(1)}_k(x;q)\ z^k\ .
\eqno(3.18)$$
These are $q$-analogs of the generating relation for the ordinary
Bessel functions $J_k(x)$,
$$ e^{x(z-z^{-1})/2}=\sum_{k=-\infty}^{\infty} \ J_k(x)\ z^k\ ,\eqno(3.19)$$
to which they reduce in the limit $q\rightarrow 1^-$, when $x$ is
replaced by $(1-q)\, x$.

\bigskip

\indent\item{\bf 3.2}{\bf Two-variable realization and addition formulas}
\medskip

We will now discuss realizations of ${\cal E}_q(2)$
on a space of functions of two complex variables, $x$ and $y$.
We shall look for a realization of $P_\pm$ and $J$ in
terms of $q$-difference operators acting on the space generated by basis
vectors of the form $f_m(x,y)=y^m\, F_m(x)$, $m=m_0+n$, $n\in {\bf Z}$,
such that (3.8) are satisfied. The constant $\omega$ is nonessential,
since it can be changed by rescaling the generators $P_+$ and $P_-$;
for simplicity, we shall henceforth set $\omega=1$.
With the help of the relations (3.6), it is easy to check that such a
realization is provided by the operators$^{12}$
$$\eqalignno{&P_+=2y\left[-\Bigl(D_x^+ -{x\over4}\Bigr)+{y\over x}\, D_y^+
\right]\, (1+T_y)^{-1}\ ,&(3.20a)\cr
             &P_-=2y^{-1}\left[\Bigl(D_x^+ -{x\over4}\Bigr)\,T_y+{y\over x}\,
D_y^+\right]\, (1+T_y)^{-1}\ ,&(3.20b)\cr
             &J=y{\partial\over\partial y}\ ,&(3.20c)\cr}$$
and the basis functions
$$ f_m(x,y)=y^m\, J^{(1)}_m(x;q)\ ,\eqno(3.21)$$
with $m=m_0+n$, $n\in{\bf Z}$.
Note that the equation $P_+\, P_-\ f_m=\omega^2\, f_m$ is now a second-order
difference equation for $J^{(1)}_m(x;q)$:
$$\eqalign{\Biggl\{\Bigl[-\Bigl(D_x^+ -{x\over4}\Bigr)
+{(1-q^{m-1})\over x}\Bigr]\,
\Bigl[\Bigl(D_x^+ -{x\over4}\Bigr)q^m+&{(1-q^m)\over x}\Bigr]\cr
&-{(1+q^m)(1+q^{m-1})\over4}\Biggr\}\,J^{(1)}_m(x;q)=\ 0\ .\cr}\eqno(3.22)$$
This is the $q$-analog of the Bessel differential equation, to which
it reduces in the limit $q\rightarrow 1^-$, provided $x$ is replaced by
$(1-q)x$.

Since we have a realization of the representation $Q(1,m_0)$,
one gets from (3.14), with $m=m_0+n$,
$$ U(\alpha,\beta,\gamma)\ f_m(x,y)=\sum_{k=-\infty}^\infty\,
U_{k\, m-m_0}(\alpha,\beta,\gamma)\ f_{m_0+k}(x,y)\ ,\eqno(3.23)$$
where the model independent matrix elements $U_{kn}(\alpha,\beta,\gamma)$
are still given by the formulas (3.15) or (3.16), with $\omega=1$.
To get addition formulas for $q$-Bessel functions, one now needs to
evaluate explicitly the l.h.s. of (3.23), {\it i.e.} to compute
directly the action of $U(\alpha,\beta,\gamma)$ on the basis functions (3.21),
when $P_\pm$ and $J$ are realized as in (3.20).
This can be done thanks to the $q$-binomial theorem (2.6) and various
identities between $q$-shifted factorials.
In the end, one arrives at the following summation formulas:$^{12}$
$$\left({x\over2}\right)^m\,
{(-2\beta/x;q)_m\over (q;q)_m}\ {}_2\phi_1\Bigl(0,-{2\beta\over x} q^m;
q^{m+1};q,-{x^2\over4}\Bigr)=\sum_{l=0}^\infty\ q^{l(l-1)/2}\,
{\beta^l\over (q;q)_l}\ J^{(1)}_{m-l}(x;q)\ ,\eqno(3.24)$$
when $\alpha=\gamma=\,0$,
$$\left({x\over2}\right)^m\,
{1\over (q;q)_m}\ {}_2\phi_1\Bigl({2\alpha \over x},0;q^{m+1};q,
-{x^2\over4}\Bigr)=\sum_{l=0}^\infty\ q^{l(l-1)/2}\,
{\alpha^l\over (q;q)_l}\ J^{(1)}_{m+l}(x;q)\ ,\eqno(3.25)$$
when $\beta=\gamma=\,0$, and for $\alpha\neq 0$ and $\beta=-q\alpha$
$$\eqalign{\left({x\over2}\right)^m\,
{(zq/xy;q)_m \over (q;q)_m}\ {}_2\phi_1\Bigl({yz\over x},&
{z\over xy}q^{m+1};q^{m+1};q,-{x^2\over4}\Bigr)\cr
&=\sum_{l=-\infty}^\infty\ q^{l(l-1)/2}\, y^l\
J^{(2)}_l(z;q)\ J^{(1)}_{m+l}(x;q)\ ,\cr}\eqno(3.26)$$
where we have set $2\alpha=z$.

The formulas (3.24) and (3.25) are the $q$-generalizations of the Lommel
summation theorems for ordinary Bessel functions$^{1,17}$
$$\eqalignno{&\Bigl(1+{2\beta\over x}\Bigr)^{m/2}\
J_m\left(x\, \Bigl(1+{2\beta\over x}\Bigr)^{1/2}\right)=\sum_{l=0}^\infty\
{\beta^l\over l!}\ J_{m-l}(x)\ ,\quad \Bigl|{2\beta\over x}\Bigr|<1\ ,
&(3.27a)\cr
           &\Bigl(1-{2\alpha\over x}\Bigr)^{-m/2}\
J_m\left(x\, \Bigl(1-{2\alpha\over x}\Bigr)^{1/2}\right)=\sum_{l=0}^\infty\
{\alpha^l\over l!}\ J_{m+l}(x)\ ,\quad \Bigl|{2\alpha\over x}\Bigr|<1\ ,
&(3.27b)\cr}$$
while (3.26) is a $q$-analog of the Graf summation theorem,$^{1,17}$
$$\eqalign{\Bigl(1-{z\over xy}\Bigr)^{m/2}\,&
\Bigl(1-{yz\over x}\Bigr)^{-m/2}\
J_m\left(x\, \Bigl(1-{z\over xy}\Bigr)^{1/2}\,
\Bigl(1-{yz\over x}\Bigr)^{1/2}\right)\cr
&\quad =\sum_{l=-\infty}^\infty\ y^l\ J_l(z)\ J_{m+l}(x)\ ,\quad
\Bigl|{z\over xy}\Bigr|<1\, ,\ \Bigl|{yz\over x}\Bigr|<1\ .\cr}
\eqno(3.28)$$
%

\vfill\eject

{\bf 4. q-OSCILLATORS AND BASIC SPECIAL FUNCTIONS}
\bigskip
\indent\item{\bf 4.1}{\bf One-variable model and q-Laguerre functions and
polynomials}
\medskip

The $q$-oscillator algebra is generated by three elements $A$, $A^\dagger$
and $N$ satisfying the defining relations$^8$
$$ [N,A]=-A\qquad\qquad [N,A^\dagger]=\, A^\dagger\ ,\eqno(4.1a)$$
$$ A A^\dagger-q\, A^\dagger A=\, 1\ .\eqno(4.1b)$$
By introducing the redefined generators
$$  a=q^{-{N\over4}}\ A\qquad\qquad a^\dagger=q^{{1-N\over4}}\ A^\dagger\ ,
\eqno(4.2)$$
the algebra (4.1) becomes
$$ \eqalign{&[N,a]=-a\cr
            &a\, a^\dagger-q^{1\over2}a^\dagger\, a=\, q^{-{N\over2}}\cr}
   \qquad\qquad
   \eqalign{&[N,a^\dagger]=\, a^\dagger\cr
            &a\, a^\dagger-q^{-{1\over2}}a^\dagger\, a=\, q^{N\over2}\ .\cr}
   \eqno(4.3)$$
This is the form in which the defining relations of the $q$-oscillator algebra
are more often presented.$^{18-20}$

In the limit $q\rightarrow 1^-$, (4.1) (and (4.3)) reduce to the canonical
commutation relations of the harmonic oscillator annihilation,
creation and number operators.
This algebra is known to have representations in which the number
operator is unbounded, or bounded from either below or above.$^1$
In this Section we shall construct $q$-deformations of these
representations;$^{11}$
they will be denoted $R_q(\omega,m_0)$, $R_q^\uparrow(\omega)$ and
$R_q^\downarrow(\omega)$.

We shall start with the representation $R_q(\omega,m_0)$ and take the following
realization of the generators $A$, $A^\dagger$ and $N$, in the space
$\cal H$ of all finite linear combinations of the monomials $z^n$,
$z\in {\bf C}$, $n\in {\bf Z}$,
$$ A={1\over 1-q} D^+_z+ {\omega+m_0\over z}\, T_q\ ,
\qquad\quad A^\dagger= z\ ,\qquad\quad N=m_0+z{d\over dz}\ ;\eqno(4.4)$$
$\omega$ and $m_0$ are complex constants ,
such that $0\leq \Re e\, m_0<1$, and $\omega+m_0$ is not an integer.
On the space $\cal H$ we shall again take as basis vectors the functions
$f_m=z^n$, with $m=m_0+n$ for all $n\in {\bf Z}$. For convenience, we set
$$ m_0+\omega={1-q^\rho\over 1-q^{\phantom{\rho}}}\ ;$$
the action of the generators on the basis vectors is then easily obtained,
$$ \eqalign{&A\, f_m=\, {1-q^{m-m_0+\rho}\over 1-q}\ f_{m-1}\ ,\cr
            &A^\dagger\, f_m=\, f_{m+1}\ ,\cr
            &N\, f_m= m f_m\ .\cr}\eqno(4.5)$$

As in the previous section we introduce the operator
$$ U(\alpha, \beta,\gamma)=\, E_q\left(\alpha(1-q)A^\dagger\right)\,
E_q\left(\beta(1-q)A\right)\, E_q\left(\gamma(1-q)N\right)\ .\eqno(4.6)$$
The matrix elements $U_{kn}(\alpha,\beta,\gamma)$ of the operator (4.6) are
defined as in (3.14). In the representation $R_q(\omega,m_0)$ they are found
to take the following form
$$U_{kn}(\alpha,\beta,\gamma)=E_q\bigl(\gamma(1-q)(m_0+n)\bigr)\
q^{{1\over2}(n-k)(n-k-1)}\, \beta^{n-k}\,
L_{\rho+k}^{(n-k)}\left(-{\alpha\beta\over q};q\right)\ .\eqno(4.7)$$
where $L_\nu^{(\lambda)}(x;q)$ stands for the $q$-Laguerre functions
$$ L_\nu^{(\lambda)}(x;q)={(q^{\lambda+1};q)_\nu\over (q;q)_\nu}\
{}_1\phi_1\Bigl(q^{-\nu};q^{\lambda+1};q;-(1-q)\, q^{\lambda+\nu+1}\,x\Bigr)
\ .\eqno(4.8)$$

A generating relation for these functions can now be obtained.
Direct evaluation shows that
$$U(\alpha,\beta,0)\, z^n=E_q\bigl(\alpha\, (1-q)\, z\bigr)\
z^n\, \bigl(-{\beta\over z};q\bigr)_{\rho+n}\ .\eqno(4.9)$$
Inserting (4.9) and (4.7) in (3.14), and taking
$n=0$, $\beta=-q$, $\gamma=\,0$ and $t=-1/z$, one obtains
$$E_q\bigl(-(1-q)\, \alpha/t\bigr)\, (-qt;q)_\rho=
\sum_{k=-\infty}^\infty\, q^{{1\over2}k(k+1)}\, t^k\
L^{(k)}_{\rho-k}(\alpha;q)\ .\eqno(4.10)$$
This is the $q$-analog of the relation$^{15,1}$
$$e^{-\alpha/t}\, (1+t)^\rho=\sum_{k=-\infty}^\infty\, t^k\
L^{(k)}_{\rho-k}(\alpha)\ ,\eqno(4.11)$$
for the ordinary Laguerre functions, to which it reduces in the limit
$q\rightarrow 1^-$.

The representation $R_q^\downarrow(\omega)$ can be realized
on the space ${\cal H}^{(+)}$ of all finite linear combinations
of the functions $z^n$, $z\in {\bf C}$, $ n\in{\bf Z}^+$, by taking
$$ A=z\ ,\qquad\quad A^\dagger={1\over 1-q}D^-_z\ ,\qquad\quad
N=-\omega-z{d\over dz}\ ,\eqno(4.12)$$
with $\omega$ a complex parameter. As basis vectors in ${\cal H}^{(+)}$
we now choose the functions $f_m=z^n$, with $m=-\omega-n$, $n\geq 0$.
The matrix elements of the operator $U(\alpha,\beta,\gamma)$, defined now by
$$ U(\alpha,\beta,\gamma)\ z^n=\sum_{k=0}^\infty\,
U_{kn}(\alpha,\beta,\gamma)\ z^k\ ,\eqno(4.13)$$
are again given in terms of the $q$-Laguerre functions.
Note that the $q$-Laguerre polynomials $L^{(\lambda)}_n(x;q)$, with $n$
integer, do not occur as matrix elements of $R_q(\omega,m_0)$ nor
$R_q^\downarrow(\omega)$. They arise
instead in the matrix elements of the representation
$R^\uparrow_q(\omega)$, to which we now come.

The representation $R^\uparrow_q(\omega)$ can be formally obtained from
the representation $R_q(\omega,m_0)$ by letting
$m_0=-\omega$ ({\it i.e.} $\rho=\,0$).
The generators of the $q$-oscillator algebra are now realized
on the space ${\cal H}^{(+)}$ as,
$$ A={1\over1-q}\, D^+_z\ ,\qquad\quad A^\dagger=z\ ,\qquad\quad
N=-\omega+z{d\over dz}\ .\eqno(4.14)$$
The basis vectors $f_m$ in
${\cal H}^{(+)}$ are defined by $f_m=z^n$, with $m=-\omega+n$. The action
of the generators (4.14) on them is as in (4.5), with $\rho-m_0$ replaced
by $\omega$.

The matrix elements of the operator $U(\alpha,\beta,\gamma)$, defined as
in (4.13), are found$^8$ to be expressible in terms of the $q$-Laguerre
polynomials,$^{21,4}$
$$ L_k^{(\lambda)}(x;q)={(q^{\lambda+1};q)_k\over(q;q)_k}\, \sum_{l=0}^k\,
{(q^{-k};q)_l\, q^{l(l-1)\over2}\,(1-q)^l\, \bigl(q^{k+\lambda+1}
x\bigr)^l\over
(q^{\lambda+1};q)_l\, (q;q)_l}\ ,\eqno(4.15)$$
as follows
$$ U_{kn}(\alpha,\beta,\gamma)=E_q\bigl(\gamma(1-q)\, (n-\omega)\bigr)\,
q^{{1\over2}(n-k)(n-k-1)}\, \beta^{n-k}\ L_k^{(n-k)}\bigl(-{\alpha\beta\over
q};
q\bigr)\ .\eqno(4.16)$$
Inserting this expression back into (4.13) and using (4.9) with $\rho=\, 0$,
one gets the following generating function for
the $q$-Laguerre polynomials:
$$ E_q\left(-\alpha(1-q)z\right)\, \bigl(-{q\over z};q\bigr)_n\, z^n=
\sum_{k=0}^\infty\, q^{{1\over2}(n-k)(n-k+1)}\, L_k^{(n-k)}(\alpha;q)\ z^k\ .
\eqno(4.17)$$
This is the $q$-analog of the relation$^{15,1}$
$$ e^{-\alpha z}\, (1+z)^n=\sum_{k=0}^\infty\, L_k^{(n-k)}(\alpha)\, z^k\ ,
\eqno(4.18)$$
for ordinary Laguerre polynomials, to which (4.17) reduces in the limit
$q\rightarrow 1^-$.

\bigskip

\indent\item{\bf 4.2}{\bf q-Oscillators and q-Hermite polynomials}
\medskip

There is another realization of the representation $R^\uparrow_q(0)$ on
functions of the complex variable $w$ which allows to relate the
$q$-oscillator algebra to $q$-Hermite polynomials. It is defined by
taking$^{8,19}$
$$ A={1\over 1-q}\, {1\over w}\, \left(1-\sqrt w\, T_w\right)\qquad\qquad
   A^\dagger=\, w\, \left(1-\sqrt{q\over w}\, T_w\right)\ ,\eqno(4.19a)$$
$$ N=\, {\ln\bigl[1-(1-q) A^\dagger A\bigr]\over \ln q}\ .\eqno(4.19b)$$
The basis functions $f_n(w)$ are here expressed in terms of the $q$-Hermite
polynomials$^{22-24}$
$$ H_n(w;q)=\, \left(w+T_w\right)^n\cdot\, 1=
\sum_{k=0}^n\, \left[{n\atop k}\right]_q\ w^k\ ,\eqno(4.20)$$
as follows:
$$  f_n(w)= \bigl(-\sqrt q\bigr)^n\,
f_0(w)\ H_n\bigl(-{w\over\sqrt q};q\bigr)\ ,\eqno(4.21)$$
with
$$ f_0(w)=\left[\sum_{k=-\infty}^\infty\, q^{k(k-1)/2}\,
w^k\right]^{1\over2}\ .\eqno(4.22)$$
The transformation properties of the basis vectors under
$U(\alpha,\beta,\gamma)$ are model independent and now provide
a relation between the $q$-Laguerre and the $q$-Hermite polynomials.
After direct evaluation of the l.h.s. of (4.13) in the realization (4.19)
and some simplifications, one obtains$^8$
$$\eqalign{ {(-w)}^n\, E_q\Bigl(& (1-q){\alpha}\,({w}+
T_{w})\Bigr)\,\cdot\, 1\cr
&\equiv {(-w)}^n\, \sum_{k=0}^\infty\, q^{k(k-1)\over2}\,
{(1-q)^k\over (q;q)_k}\
{\alpha}^k\, H_k(w;q)\cr
&=\sum_{k=0}^\infty\, (-1)^k q^{{1\over2}[k(k-n+1)-n(n+k+1)]}\
L_k^{(n-k)}\bigl(q^{-(n+1)}\alpha;q\bigr)\, H_k\bigl(q^n w;q
\bigr)\ .\cr}\eqno(4.23)$$
It should be noted that
$$\lim_{q\rightarrow 1^-} H_n(w;q)=\, (w+1)^n\ .\eqno(4.24)$$
Substituting $z$ for $-(w+1)$, it easily seen that (4.23) goes into
(4.18) in the limit $q\rightarrow 1^-$.

\bigskip

\indent\item{\bf 4.3}{\bf The metaplectic representation of
su$_{\hbox{\bf q}}$(1,1) and the q-Gegenbauer polynomials}
\medskip

The metaplectic representation of $su_q(1,1)$ is defined from the
representation (4.14) of the $q$-oscillator algebra by taking
the generators $K_\pm$ and $K_0$ of $su_q(1,1)$ to be given by$^9$

$$ K_+={1\over [2]_{q^{1/2}}} \bigl(A^\dagger\bigr)^2\qquad
   K_-={1\over [2]_{q^{1/2}}} A^2\qquad
   K_0={1\over2}\bigl(N+{1\over2}\bigr)\ .\eqno(4.25)$$
The defining relations
$$ \eqalign{&K_-\, K_+-q^2\, K_+\, K_-=q^{2K_0}\, [2K_0]_q\cr
            &[K_0,K_\pm]=\pm\, K_\pm\ ,\cr}\eqno(4.26)$$
are thus realized. The following redefinition of the generators
$\tilde J_\pm=\pm q^{-K_0+{1\over2}\pm{1\over2}}\ K_\pm$, $\tilde J_3=K_0$
casts the defining relations in the standard form$^{6,7}$
$$\eqalign{&[\tilde J_+,\tilde J_-]={q^{2\tilde J_3}-q^{-2\tilde J_3}
\over q-q^{-1}}\cr
            &[\tilde J_3,\tilde J_\pm]=\pm \tilde J_\pm\ .\cr}\eqno(4.27)$$
This representation
decomposes into two irreducible components, with the invariant
subspaces ${\cal H}^{(e)}$ and ${\cal H}^{(o)}$
formed out of $N$-eigenstates with even and odd eigenvalues,
respectively.

Let
$$ U(\alpha, \beta,\gamma)=\, E_{q^2}\left(\alpha(1-q^2)[2]_{q^{1/2}}
K_+\right)\,
E_{q^2}\left(\beta(1-q^2)[2]_{q^{1/2}} K_-\right)\,
E_{q^2}\left(\gamma(1-q^2)2K_0\right)
\ ;\eqno(4.28)$$
its matrix elements in the
spaces ${\cal H}^{(e)}$ and ${\cal H}^{(o)}$ are defined by
$$ \eqalignno{&U(\alpha,\beta,\gamma)\ z^{2n}=\sum_{k=0}^\infty\,
U_{kn}^{(e)}(\alpha,\beta,\gamma)\ z^{2k} &(4.29a)\cr
&U(\alpha,\beta,\gamma)\ z^{2n+1}=\sum_{k=0}^\infty\,
U_{kn}^{(o)}(\alpha,\beta,\gamma)\ z^{2k+1}\ . &(4.29b)\cr}$$
They are evaluated to be
$$ \eqalign{U_{kn}^{(e)}(\alpha,\beta,\gamma)=E_{q^2}&
\left(\gamma (1-q^2)\bigl(2n+{1\over2}
\bigr)\right)\,q^{(k+n)^2+k-n} {(q;q)_{2n}\over(q^2;q^2)_{n+k}}\,
\alpha^k\beta^n\cr
&\times  (1+q)^{k+n}\,(1-q)^{k-n}\
{\cal C}_{2k}^{(-n-k)}\Bigl(q;{q\over(1+q){\sqrt{\alpha\beta}}}\Bigr)\ ,
\cr}\eqno(4.30)$$
\null
$$ \eqalign{U_{kn}^{(o)}(\alpha,\beta,\gamma)&=-E_{q^2}
\left(\gamma (1-q^2)\bigl(2n+{3\over2}
\bigr)\right)\,q^{(k+n+1)^2+k-n} {(q;q)_{2n+1}\over(q^2;q^2)_{n+k+1}}\,
\alpha^k\beta^n\cr
&\times\sqrt{\alpha\beta}\, (1+q)^{k+n+1}\,(1-q)^{k-n}\
{\cal C}_{2k+1}^{(-n-k-1)}
\Bigl(q;{q\over(1+q){\sqrt{\alpha\beta}}}\Bigr)\ ,
\cr}\eqno(4.31)$$
where ${\cal C}^{(\lambda)}_k$ are the following $q$-generalizations of the
Gegenbauer polynomials
$$ {\cal C}^{(\lambda)}_k(q;z)\equiv{\cal C}_k(q;q^{2\lambda};z)
=\sum_{l=0}^{\llb k/2\rrb}
{(-1)^l q^{l(l-1)} (q^{2\lambda};
q^2)_{k-l}\over (q^2;q^2)_l\, (q;q)_{k-2l}}\, z^{k-2l}\ ,\eqno(4.32)$$
with $\llb x\rrb$ the integer part of $x$. They obey the three-term
recurrence relation
$$ (1-q^{k+1})\,{\cal C}_{k+1}^{(\lambda)}(q;z)-(1-q^{2(\lambda+k)})\,z\,
{\cal C}_k^{(\lambda)}(q;z)+q^{k-1}(1-q^{2\lambda+k-1})\,
{\cal C}_{k-1}^{(\lambda)}(q;z)=\, 0\ ,\eqno(4.33)$$
and thus form an orthogonal set.
Introducing the discrete $q$-Hermite polynomials$^4$
$$\eqalign{h_k(q;z)=&\, (q;q)_k\ {\cal C}_k(q;0;z)\cr
=&\sum_{l=0}^{\llb k/2\rrb}
{(-1)^l q^{l(l-1)}\, (q;q)_k \over (q^2;q^2)_l\, (q;q)_{k-2l}}\, z^{k-2l}\
,\cr}
\eqno(4.34)$$
and proceeding as before, one gets the following generating functions
for the $q$-Gegenbauer polynomials:$^9$
$$\eqalignno{&E_{q^2}\Bigl(-{w^2\over q^2}\Bigr)\,
h_{2n}\Bigl(q; {wx\over q^2}\Bigr)\cr
&\hskip 1.5cm
=\sum_{k=0}^\infty (-1)^{n+k}\, q^{(n+k)(n+k-1)}\,
{(q;q)_{2n}\over(q^2;q^2)_{n+k}}
\, {\cal C}_{2k}^{(-n-k)}(q;x)\ w^{2k}\ ,&(4.35a)\cr
&\null\cr
&E_{q^2}\Bigl(-{w^2\over q^2}\Bigr)\, h_{2n+1}\Bigl(q;{wx\over q^2}\Bigr)\cr
&\hskip 1cm
=\sum_{k=0}^\infty (-1)^{n+k+1}\, q^{(n+k+1)(n+k)}\,
{(q;q)_{2n+1}\over(q^2;q^2)_{n+k+1}}
\, {\cal C}_{2k+1}^{(-n-k-1)}(q;x)\ w^{2k+1}\ .&(4.35b)\cr}$$
In the limit $q\rightarrow 1^-$, ${\cal C}^{(\lambda)}_k(q;z)$ become
the ordinary Gegenbauer polynomials ${\cal C}^{(\lambda)}_k(z)$ and
$(1+q)^{k/2}(1-q)^{-k/2}\, h_k\bigl(q;\sqrt{1-q^2}\, z\bigr)$
the usual Hermite polynomials $H_k(z)$;
in this limit the equations (4.35) tend to
$$\eqalignno{& e^{-w^2}\, H_{2n}(wx)=\sum_{k=0}^\infty (-1)^{n+k}\,
{(2n)!\over (n+k)!}\, {\cal C}^{(-n-k)}_{2k}(x)\, w^{2k}\ ,&(4.36a)\cr
&\null\cr
& e^{-w^2}\, H_{2n+1}(wx)=\sum_{k=0}^\infty (-1)^{n+k+1}\,
{(2n+1)!\over (n+k+1)!}\, {\cal C}^{(-n-k-1)}_{2k+1}(x)\, w^{2k+1}\ .
&(4.36b)\cr}$$
These relations express the usual Hermite polynomials in terms of the
usual Gegenbauer polynomials.

\vfill\eject

{\bf 5. THE QUANTUM ALGEBRA {\titlefont sl}$_{\hbox{q}}$(2) AND
q-HYPERGEOMETRIC\hfill\break
\indent\phantom{\bf 5. }FUNCTIONS}
\bigskip
\indent\item{\bf 5.1}{\bf Realizations in terms of first-order q-difference
operators}
\medskip

The quantum algebra $sl_q(2)$ is generated by three elements
$J_+$, $J_-$ and $J_3$ satisfying the defining relations
$$ \eqalign{&J_+\, J_- -q^{-1}\, J_-\, J_+=
{1-q^{2 J_3}\over 1-q^{\phantom{2J_3}}}\cr
            &[J_3,J_\pm]=\pm J_\pm\ .\cr}\eqno(5.1)$$
In the limit $q\rightarrow 1^-$, the relations (5.1) reduce to the
Lie brackets of the ordinary algebra $sl(2)$.
If we redefine the generators according to $\tilde J_\pm=
q^{-{1\over2}(J_3\mp{1\over2})}\, J_\pm$, $\tilde J_3= J_3$, the algebra
(5.1) takes the form given in (4.27) with $q$ replaced by $q^{1/2}$.

The ordinary algebra $sl(2)$ is known to have representations
that are unbounded, bounded from either below or above and finite
dimensional.$^1$ In the following we shall consider the $q$-analogs of
these representations,$^{11}$ and
call them $D_q(u,m_0)$, $D_q^\uparrow(j)$, $D_q^\downarrow(j)$
and $D_q(2j)$, respectively.

The representation $D_q(u,m_0)$ is characterized by two complex constants
$u$ and $m_0$ such that neither $m_0+u$ nor $m_0-u$ is an integer, and
$0\leq \Re e\, m_0<1$. On the space $\cal H$ of all finite linear combinations
of the functions $z^n$, $z\in {\bf C}$, $n\in {\bf Z}$,
the generators $J_\pm$ and $J_3$ are realized as (see also Ref.[25]),
$$\eqalignno{&J_+=q^{{1\over2}(m_0-u+1)}\left[ {z^2\over1-q} D^+_z
-{1-q^{u-m_0}\over 1-q}\, z\right]\ ,&(5.2a)\cr
             &J_-=-q^{{1\over2}(m_0-u+1)}\left[ {1\over1-q} D^+_z
+{1-q^{u+m_0}\over 1-q}\,{1\over z}\, T_q\right]\ ,&(5.2b)\cr
             &J_3=m_0+z{d\over dz}\ .&(5.2c)\cr}$$
The basis vectors $f_m$ in $\cal H$ are still defined by $f_m=z^n$ for
$m=m_0+n$ and all integers $n$. Thus,
$$\eqalign{&J_+\, f_m=q^{{1\over2}(u-m_0+1)}\, {1-q^{m-u}\over 1-q}\
f_{m+1}\ ,\cr
           &J_-\, f_m=-q^{{1\over2}(m_0-u+1)}\, {1-q^{m+u}\over 1-q}\
f_{m-1}\ ,\cr
           &J_3\, f_m= m\, f_m\ .\cr}\eqno(5.3)$$

As in the previous sections, we introduce the operator
$$ U(\alpha, \beta,\gamma)=\, E_q\left(\alpha(1-q)J_+\right)\,
E_q\left(\beta(1-q)J_-\right)\, E_q\left(\gamma(1-q)J_3\right)\ ,\eqno(5.4)$$
acting on the space $\cal H$ through the realization (5.2),
and define its matrix elements again as in (3.14).
The elements $U_{kn}(\alpha,\beta,\gamma)$ are expressed in terms of the
basic hypergeometric function $_2\phi_1$:
$$\eqalign{U_{kn}(\alpha,\beta,&\gamma)=E_q\left(\gamma\, (1-q)
\Bigl(n+{1\over2}(s-t)\Bigr)\right)\
\Bigl(q^{s+n+{1\over2}(1-t)}\, \beta\Bigr)^{n-k}\cr
&\times\ {(q^{-s-n};q)_{n-k}\over (q;q)_{n-k}}\,
{}_2\phi_1\Bigl(q^{-s-k},q^{t-k+1};q^{n-k+1};q;- q^{s-t+n+k}\,
\alpha\beta\Bigr)\ ,\quad
{\rm if}\ k\leq n\ ,\cr}\eqno(5.5a)$$
\null
$$\eqalign{U_{kn}(\alpha,\beta,&\gamma)=E_q\left(\gamma\, (1-q)
\Bigl(n+{1\over2}(s-t)\Bigr)\right)\
\Bigl(-q^{k-1+{1\over2}(1-t)}\, \alpha\Bigr)^{k-n}\cr
&\times\ {(q^{t-k+1};q)_{k-n}\over (q;q)_{k-n}}\,
{}_2\phi_1\Bigl(q^{-s-n},q^{t-n+1};q^{k-n+1};q;- q^{s-t+n+k}\,
\alpha\beta\Bigr)\ ,\quad
{\rm if}\ k\geq n\ ,\cr}\eqno(5.5b)$$
where $s=u+m_0$ and $t=u-m_0$. Note that neither $s$ nor $t$ is an integer.
Thanks to the limiting relation (2.16), either one of the
two expressions for $U_{kn}(\alpha,\beta,\gamma)$ remain valid over the
whole range of the indices.
Using these results the following identity can be obtained from (3.14)
$$\eqalign{\sum_{r=0}^\infty\, {(q^{-s};q)_r\over (q;q)_r}\, \beta^r\
{}_1\phi_1&\Bigl(q^{-r-t};-;q;q^{t+1}\, {x\over\beta}\Bigr)\cr
&=\sum_{k=-\infty}^{\infty}\, {(q^{-s};q)_k\over (q;q)_k}\, \beta^k\
{}_2\phi_1\Bigl(q^{k-s},q^{t+k+1};q^{k+1};q;q^{-k}\, x\Bigr)\ ,\cr}
\eqno(5.6)$$
which can be considered as a generating function for $_2\phi_1$.

The representation $D_q^\uparrow(j)$ can be obtained from the one
just discussed by setting $m_0=-u\equiv-j$, with $2j$ not a non-negative
integer.
The generators $J_\pm$ and $J_3$ act now on the space ${\cal H}^{(+)}$
of all finite linear combinations
of the functions $z^n$, $z\in {\bf C}$, $n\in {\bf Z}^+$.
The analysis of this case leads to the following generating
function for the terminating series $_2\phi_1$ as a finite combination of
functions $_1\phi_1$
$$\eqalign{\sum_{m=0}^n(-1)^n\, q^{{1\over2}(m(m+1)-n(n+1))}\,&
\left[{n\atop m}\right]_q
(-z)^l\ {}_1\phi_1\Bigl(q^{m-2j};-;q;q^{2j+1}\, xz\Bigr)\cr
&=\sum_{k=0}^\infty\, {(q^{-n};q)_{n-k}\over (q;q)_{n-k}}\, z^k\
{}_2\phi_1\Bigl(q^{-k},q^{2j-k+1};q^{n-k+1};q;q^k\, x\Bigr)\ .}\eqno(5.7)$$

The representation $D_q^\downarrow(j)$ of $sl_q(2)$ can be formally obtained
from $D_q^\uparrow(j)$ by letting $J_+\rightarrow q^{-(J_3+1/2)}\, J_-$,
$J_-\rightarrow q^{-(J_3-1/2)}\, J_+$ and $J_3\rightarrow -J_3$.

We now come to the finite-dimensional representation $D_q(2j)$,
which is defined for $2j$ a nonnegative integer. The generators $J_\pm$
and $J_3$ are realized as in $D_q^\uparrow(j)$, but they now act on the
finite-dimensional space ${\cal H}^{(j)}$ of all linear combinations of
the functions $z^n$, $z\in{\bf C}$, $n=0,1,\ldots,2j$.
For the basis vectors $f_m$, we take
$f_m=z^n$, with $m=-j+n$, $0\leq n\leq 2j$.

The matrix elements of the operator $U(\alpha,\beta,\gamma)$ are
now defined by
$$ U(\alpha,\beta,\gamma)\ z^n=\sum_{k=0}^{2j}\,
U_{kn}(\alpha,\beta,\gamma)\ z^k\ ,\eqno(5.8)$$
and can be expressed
in terms of little $q$-Jacobi polynomials,$^4$
$$ p_n(x;a;b;q)={}_2\phi_1\bigl(q^{-n},abq^{n+1};aq;q;qx\bigr)\ .\eqno(5.9)$$
Indeed, one has
$$\eqalign{U_{kn}(\alpha,\beta,\gamma)=E_q\bigl(\gamma\, (1-q)\,(n-j)\bigr)\,
&{(q^{-n};q)_{n-k}\over (q;q)_{n-k}}\,
\left(q^{n-j+1/2}\, \beta\right)^{n-k}\cr
&\times\ p_k\Bigl(-q^{n+k-2j-1}\,\alpha\beta;
q^{n-k};q^{2j-k-n};q\Bigr)\ .\cr}\eqno(5.10)$$

A straightforward computation shows that the action of $U(\alpha,\beta,0)$
on $z^n$ can also be expressed in terms of the
Stieljes-Wiegert polynomials,$^{26,27}$
$$s_n(x;q)=(-1)^n\, q^{(2n+1)/4}\, (q;q)_n^{-1/2}\
\sum_{l=0}^n\, \left[{n\atop l}\right]_q\, q^{l^2}\,
(-q^{1/2}\,x)^l\ .\eqno(5.11)$$
One finds,
$$ \eqalign{U(\alpha,\beta,0)\, z^n=\sum_{m=0}^n\,
&q^{{1\over2}\bigl((n-m)(n-m-1)-2j+m+1/2\bigr)}\,
\left[{n\atop m}\right]_q\,\bigl((q;q)_{2j-m}\bigr)^{1/2}\,\cr
&\hskip 1cm \times\ \left(-q^{-j+1/2}\beta\right)^{n-m}\, (-z)^m\
s_{2j-m}\bigl(q^{m-j-1}\, \alpha z;q\bigr)\ .\cr}\eqno(5.12)$$
Inserting (5.12) and (5.10) in (5.8), after some redefinitions
one gets
$$\eqalign{\sum_{m=0}^n & q^{-{1\over2}(n(n-m-1)+2j+1/2)}\,
\left[{n\atop m}\right]_q\, \bigl((q;q)_{2j-m}\bigr)^{1/2}\
\bigl(-q^{{1\over2}(m-n)}\, z\bigr)^{m}\
s_{2j-m}\bigl(q^{m-1/2}\, wz;q\bigr)\cr
&\hskip 2cm =\sum_{k=0}^{2j}\, q^{{1\over2}(k(k-1)-n(n-1))}\,
\left[{n\atop k}\right]_q\, z^k\
p_k\Bigl(q^k\, w;q^{n-k};q^{2j-k-n};q\Bigr)\ .\cr}\eqno(5.13)$$
This generating formula can also be equivalently rewritten in terms
of the Wiegert-Szeg\"o polynomials,$^{22-24}$
$$ G_n(x;q)=\sum_{l=0}^n\, \left[{n\atop l}\right]_q\,
q^{l(l-n)}\ x^l\ .\eqno(5.14)$$
One easily proves that
$$\eqalign{\sum_{m=0}^n\, &q^{-{1\over2}(n-m)(n+m-1)}\,
\left[{n\atop m}\right]_q\, z^m\, G_{2j-m}(-q^{2j}\,wz)\cr
&\hskip 1cm =\sum_{k=0}^{2j}\, q^{{1\over2}(k(k-1)-n(n-1))}\,
\left[{n\atop k}\right]_q\, z^k\
p_k\Bigl(q^k\, w;q^{n-k};q^{2j-k-n};q\Bigr)\ .\cr}\eqno(5.15)$$

\bigskip

\indent\item{\bf 5.2}{\bf A realization in terms of second-order q-difference
operators}
\medskip

We can also represent the generators of the quantum
algebra $sl_q(2)$ on the space $\cal H$ using quadratic
expressions in $q$-difference operators. Here it is more convenient
to use generators $K_\pm$ and $K_0$ subjected to the defining relations (4.26).
One takes$^{10}$
$$\eqalignno{K_-=&\,{1\over[2]_{q^{1/2}}}\ \left[{1\over(1-q)^2} D_z^2+
{\omega\over z^2}\, T_q\right]\ ,&(5.16a)\cr
           K_+=&\,{1\over[2]_{q^{1/2}}}\ z^2\ ,&(5.16b)\cr
           K_0=&\,{1\over2}\left(z{d\over dz}+{1\over2}\right)\ ,&(5.16c)\cr}$$
where $\omega$ is a complex parameter. This representation
decomposes into two irreducible components, with the invariant
subspaces ${\cal H}^{(e)}$ and ${\cal H}^{(o)}$
respectively spanned by the even and odd powers of $z$.

The operator $U(\alpha,\beta,\gamma)$ is as in (4.28)
and its matrix elements in the
spaces ${\cal H}^{(e)}$ and ${\cal H}^{(o)}$ are defined by
$$ \eqalignno{&U(\alpha,\beta,\gamma)\ z^{2n}=\sum_{k=-\infty}^\infty\,
U_{kn}^{(e)}(\alpha,\beta,\gamma)\ z^{2k}\ , &(5.17a)\cr
&U(\alpha,\beta,\gamma)\ z^{2n+1}=\sum_{k=-\infty}^\infty\,
U_{kn}^{(o)}(\alpha,\beta,\gamma)\ z^{2k+1}\ . &(5.17b)\cr}$$
With
$$\omega={1+q^{-1}-q^\rho-q^{-(\rho+1)}\over 1-q^2}\ ,$$
one obtains for these elements the following expressions
$$ \eqalign{U_{kn}^{(e)}(\alpha,\beta,\gamma)=E_{q^2}&
\left(\gamma (1-q^2)\bigl(2n+{1\over2}\bigr)\right)\,
{q^{(k-n)(k-n-1)}\over (q^2;q^2)_{k-n}}\,\left((1-q^2)\, \alpha\right)^{k-n}
\cr
&\quad\times\ {}_2\phi_1\Bigl(q^{-\rho-2n},q^{\rho-2n+1};q^{2k-2n+2};q^2;
 \alpha\beta\, (1+q)^2\, q^{2n+2k-1}\Bigr)\ ,
\cr}\eqno(5.18)$$
\null
$$ \eqalign{U_{kn}^{(o)}(\alpha,\beta,\gamma)=E_{q^2}&
\left(\gamma (1-q^2)\bigl(2n+{3\over2}\bigr)\right)\,
{q^{(k-n)(k-n-1)}\over (q^2;q^2)_{k-n}} \left((1-q^2)\, \alpha\right)^{k-n}\cr
&\quad\times\ {}_2\phi_1\Bigl(q^{\rho-2n},q^{-\rho-2n-1};q^{2k-2n+2};q^2;
 \alpha\beta (1+q)^2\, q^{2n+2k+1}\Bigr)\ .
\cr}\eqno(5.19)$$
These formulas are defined for the whole range of $k$ and $n$ with the help of
the limiting relation (2.16).

The generating function which is here obtained from $(5.17a)$ reads as follows
$$\eqalign{E_{q^2}(w)\, w^n\
_2\phi_0 & \Bigl(q^{-\rho-2n},q^{\rho-2n+1};q^2;
-q^{2n}{\alpha\over w}\Bigr)\cr
&=\sum_{k=-\infty}^\infty w^k\,
{q^{(k-n)(k-n-1)}\over (q^2;q^2)_{k-n}}\
{}_2\phi_1\Bigl(q^{-\rho-2n},q^{\rho-2n+1};q^{2k-2n+2};q^2;
 \alpha\, q^{2k}\Bigr)\ .\cr}\eqno(5.20)$$
An equivalent formula results from $(5.17b)$.

\vskip 2cm

\centerline{\bf REFERENCES}
\bigskip
\item{1.} Miller, W., {\it Lie Theory and Special Functions},
(Academic Press, New York, 1968)
\smallskip
\item{2.} Vilenkin, N.Ya., {\it Special Functions and the Theory of
Group Representations}, Amer. Math. Soc. Transl. of Math.
Monographs {\bf 22}, (The American Mathematical Society, Providence, 1968)
\smallskip
\item{3.} Nikiforov, A. and Ouvarov, V., {\it El\'ements de la theorie
des fonctions sp\'eciales}, (Mir, Moscow, 1976)
\smallskip
\item{4.} Gasper, G. and Rahman, M., {\it Basic Hypergeometric Series},
(Cambridge University Press, Cambridge, 1990)
\smallskip
\item{5.} Miller, W., Lie theory and $q$-difference equations,
SIAM J. Math. Anal. {\bf 1}, 171-188 (1970)
\smallskip
\item{6.} Drinfel'd, V.G., Quantum groups, in: {\it Proceedings of the
International Congress of Mathematicians}, Berkeley (1986), vol. {\bf 1},
pp. 798-820, (The American Mathematical Society, Providence, 1987)
\smallskip
\item{7.} Jimbo, M., A $q$-difference analogue of $U(g)$ and the Yang-Baxter
equation, Lett. Math. Phys. {\bf 10}, 63-69 (1985); A $q$-analogue of
$U(gl(N+1))$, Hecke algebra and the Yang-Baxter equation, {\it ibid.} {\bf 11},
247-252 (1986)
\smallskip
\item{8.} Floreanini, R. and Vinet, L., $q$-Orthogonal polynomials
and the oscillator quantum group, Lett. Math. Phys. {\bf 22}, 45-54 (1991)
\smallskip
\item{9.} Floreanini, R. and Vinet, L., The metaplectic representation
of $su_q(1,1)$ and the $q$-Gegenbauer polynomials,
University of Montr\'eal-preprint, UdeM-LPN-TH10, 1991
\smallskip
\item{10.} Floreanini, R. and Vinet, L., $q$-Hypergeometric functions
and $sl_q(2)$, University of Montr\'eal-preprint, UdeM-LPN-TH51, 1991
\smallskip
\item{11.} Floreanini, R. and Vinet, L., Quantum algebras and $q$-special
functions, University of Montr\'eal-preprint, UdeM-LPN-TH54, 1991
\smallskip
\item{12.} Floreanini, R. and Vinet, L., Addition formulas for $q$-Bessel
functions, University of Montr\'eal-preprint, UdeM-LPN-TH60, 1991
\smallskip
\item{13.} Agarwal, A.K., Kalnins, E.G. and Miller, W., Canonical
equations and symmetry techniques for $q$-series, SIAM J. Math. Anal.
{\bf 18}, 1519-1538 (1987)
\smallskip
\item{14.} Kalnins, E.G., Manocha, H.L. and Miller, W., Models of
$q$-algebra representations: I. Tensor products of special unitary
and oscillator algebras, University of Minnesota preprint, 1991
\smallskip
\item{15.} {\it Higher Transcendental Functions}, Erd\'elyi, A., ed.,
(McGraw-Hill, New York, 1953)
\smallskip
\item{16.} Vaksman, L.L. and Korogodskii, L.I., Algebra of bounded functions
on the quantum group of the motions of the plane, and $q$-analogues of
Bessel functions, Soviet. Math. Dokl. {\bf 39}, 173-177 (1989)
\smallskip
\item{17.} Watson, G.N., {\it Theory of Bessel Functions}, (Cambridge
University Press, Cambridge, 1962)
\smallskip
\item{18.} Biedenharn, L.C., The quantum group $SU(2)_q$ and a $q$-analogue
of the boson operators, J. Phys. {\bf A22}, L873-L878 (1989)
\smallskip
\item{19.} Macfarlane, A.J., On $q$-analogues of the quantum harmonic
oscillator and the quantum group $SU(2)_q$,
J. Phys. {\bf A22}, 4581-4588 (1989)
\smallskip
\item{20.} Hayashi, T., Q-analogue of Clifford and Weyl algebras--Spinor and
oscillator representations of quantum envelopping algebras,
Comm. Math. Phys. {\bf 127}, 129-144 (1990)
\smallskip
\item{21.} Moak, D.S., The $q$-analogue of the Laguerre polynomials,
J. Math. Anal. Appl. {\bf 81}, 20-47 (1981)
\smallskip
\item{22.} Carlitz, L., Some polynomials related to theta functions,
Annali di Matematica Pura ed Applicata (4) {\bf 41}, 359-373 (1955)
\smallskip
\item{23.} Carlitz, L., Some polynomials related to theta functions,
Duke Math. J. {\bf 24}, 521-527 (1957)
\smallskip
\item{24.} Carlitz, L., Note on orthogonal polynomials related to theta
functions, Publicationes Mathematicae {\bf 5}, 222-228 (1958)
\smallskip
\item{25.} Jur\v co, B., On coherent states for the simplest quantum
groups, Lett. Math. Phys. {\bf 21}, 51-58 (1991)
\smallskip
\item{26.} Chihara, T.S., {\it An Introduction to Orthogonal Polynomials},
(Gordon and Breach, New York, 1978)
\smallskip
\item{27.} Szeg\"o, G., {\it Orthogonal Polynomials}, (American
Mathemetical Society, New York, 1959)

\bye